\documentclass[conference,letterpaper]{IEEEtran}
\IEEEoverridecommandlockouts
\usepackage{cite}
\usepackage{amsmath,amsthm,amssymb,amsfonts}
\usepackage{algorithmic}
\usepackage{graphicx}
\usepackage{epstopdf}
\usepackage{textcomp}
\usepackage{xcolor}
\usepackage{balance}
\usepackage{tikz}

\newtheorem {Proposition}{Proposition}

\newtheorem {Corollary}{Corollary}

\def\BibTeX{{\rm B\kern-.05em{\sc i\kern-.025em b}\kern-.08em
    T\kern-.1667em\lower.7ex\hbox{E}\kern-.125emX}}
\begin{document}

\title{Bit Error Rate Performance and Diversity Analysis for Mediumband Wireless Communication
}
\author{\IEEEauthorblockN{
		Dushyantha A Basnayaka \IEEEauthorrefmark{1}, and 
		Jiabin Jia \IEEEauthorrefmark{2}  
	}                                    
	\IEEEauthorblockA{\IEEEauthorrefmark{1}
		School of Electronics Engineering, Dublin City University, Collins Ave, Dublin 9, Ireland}
	\IEEEauthorblockA{\IEEEauthorrefmark{2}
		School of Engineering, The University of Edinburgh, Edinburgh, EH9 3FB.}
	E-mail: d.basnayaka@dcu.ie
}

\maketitle

\begin{abstract}
Mediumband wireless communication refers to wireless communication through a class of channels known as ``\textit{mediumband}'' that exists on the $T_mT_s$-plane. This paper, through statistical analysis and computer simulations, studies the performance limits of this class of channels in terms of uncoded bit error rate (BER) and diversity order. We show that, owing mainly to the effect of the deep fading avoidance, which is unique to the channels in the mediumband region, mediumband wireless systems, if designed judiciously, have the potential to achieve significantly superior error rate and higher order diversity even in non-line-of-sight (NLoS) propagation environments where the achievable diversity order is otherwise low.    
\end{abstract}
\vspace{-3mm}
\section{Introduction}
%
%
%
In modern digital wireless communication, the design of wireless systems is such that the effect of the environment between the transmitter (TX) and the receiver (RX) on the performance is significant. This environment is typically outside the control of the wireless systems and has a random nature. However, this conventional line of thought is now slowly changing and considerable research is ongoing to find innovative ways to harness the potential of the environment for more reliable and energy efficient wireless communication \cite{Zhang23,Mei22}.\\
\indent At the forefront of current research into harnessing the potential of the propagation environment for future wireless communication is finding effective methods to introduce foreign objects like ``\textbf{\textit{reflecting surfaces}}'' into the wireless environment \cite{Zhang23}. These reflecting surfaces have a large number of passive or active elements that steer electromagnetic signals so as to reconfigure the propagation environment. Without making such interventions to the wireless environment, in this paper, we study an emerging area of study that deals with a new class of systems that can successfully alter ``\textit{\textbf{the effect of the wireless environment}}'' without changing the environment physically. It has been shown recently that communicating in the mediumband can non-intrusively statistically alter the effects of the wireless environment increasing the reliability of wireless communication significantly \cite{Bas2023}.
%
%

%
%
%
\subsection{Mediumband Systems}
\begin{figure}[t]
	\centerline{\includegraphics*[scale=0.6]{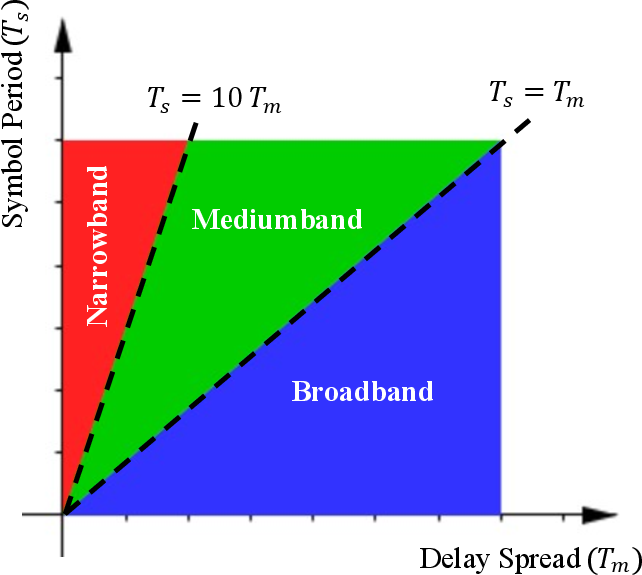}}
	\caption{Three main regions on the $T_mT_s$--plane \cite{Bas2023}.}\label{fig:fig0}
	\vspace{-5mm}
\end{figure}

As shown in Fig. \ref{fig:fig0}, mediumband systems, which occurs when $T_m \leq T_s \leq 10T_m$, is a class of systems that falls in the transitional region between narrowband and the broadband regions \cite{Bas2023}. In this regime, the effect of multipath has been shown to be two-fold. As in the narrowband case, the effect of multipath appears as a multiplicative fading factor for the desired signal. The multipath also gives rise to an additive inter-symbol-interference (ISI), which has been shown to increase as the degree of mediumband-ness increases \cite[eq. 13]{Bas2023}. However, since the symbol period can be made to be smaller than that of narrowband systems, signalling can be done at a higher rate in mediumband systems enabling significantly higher data rates.   
The percentage delay spread (PDS):
%
%
\begin{align}
	\text{Percentage Delay Spread (PDS)} &= \left(\frac{T_m}{T_s}\right) \times 100 \%,
\end{align}
is used to capture the degree of meadiumband-ness of mediumband systems as a percentage. Also, as the PDS increases, it has been shown that an effect known as ``\textit{\textbf{deep fading avoidance}}'' becomes dominant, and appears to significantly reduce the deep fading countering the adverse effects of ISI significantly.\\ 
%
%
%
\begin{figure*}
	\begin{align} \tag{7} \label{alpha:eq1_opt}
		\eta_o &= \sqrt{\left(1-\frac{\beta}{4}\right)\left[\left(\sum_{n=0}^{N-1} |\gamma_n|^2 \right) - |h_o|^2 \right] + \sum_{n=0}^{N-1} \sum_{\substack{m=0 \\ m \neq n}}^{N-1} \gamma_n \gamma_m^* R(\tau_n-\tau_m)}.
	\end{align} 
	%
	\begin{align} \tag{8} \label{auto-corr:eq1}
		R\left(\tau\right) = \left.\operatorname{sinc}\left( \frac{\tau}{T_s} \right) \frac{\cos\left( \beta \frac{\pi \tau}{T_s} \right)}{1 - \left( \frac{2 \beta \tau}{T_s} \right)^2} - \frac{\beta}{4} \operatorname{sinc}\left(\beta \frac{\tau}{T_s} \right) \frac{\cos\left( \frac{\pi \tau}{T_s} \right)}{1 - \left( \frac{\beta \tau}{T_s} \right)^2} \right.
	\end{align}
\hrule
\end{figure*}
\indent The current paper studies the effect of mediumband-ness on non-line-of-sight (NLoS) wireless communication from the perspective of uncoded bit error rate (BER). The desired fading factor in mediumband regime possesses a unique bimodal distribution (see Fig. \ref{fig:secv:fig1}) \cite{Bas2023, BasJ23}. As a result, we show in this paper that, mediumband wireless communication has the potential to achieve significantly superior uncoded error performance in NLoS fading. The study, supported by both statistical analysis and simulation, confirm the benefits of operating in the mediumband regime.

\section{Propagation Environment}
A radio frequency (RF) wireless communication system with a single TX and a single RX in a rich NLoS scattering environment is considered. As a result of multipath, the received RF signal is a mixture of signals. After RF mixing and low pass filtering, this RF signal is converted to a baseband equivalent electrical signal, and let it be defined by $r(t)$. In the absence of noise, this $r(t)$ can be expressed mathematically as \cite{Gold05}:
\begin{align}\label{eq:channel:1}
	r(t) &= \sqrt{E_s} \sum_{n=0}^{N-1} \alpha_n e^{-j\phi_n} s(t-\tau_n),
\end{align}
where $N$ is the number of multipath components, and $\tau_n$ and $\alpha_n$ are the absolute time delay and the path gain of the $n$th component respectively. Furthermore, $\phi_n=2\pi F_c\tau_n$ is known as the phase of the $n$th component, and $F_c$ here is the carrier frequency. The $s(t)$ is the baseband equivalent transmitted signal corresponding to a single frame, and $\sqrt{E_s}$ is a factor that captures the effect of transmit power. It is assumed that all the propagation parameters, i.e., $\tau_n$, $\alpha_n$, and $\phi_n$ are fixed at least within the time duration of a single frame corresponding to the case of static terminals or terminals with slow relative movement. Without loss of generality, $0$ is assumed to be the path index of the earliest path meaning $\alpha_0$ and $\tau_0$ are the path gain and absolute delay of the earliest (also the shortest) path.\\
\indent Combining the effects of square-root-raised-cosine (SRRC) transmit and receive pulse shaping filters, the baseband equivalent signal, $s(t)$ can be given by \cite{Proakis00}:
\begin{align} \label{eq1}
	s(t) &= \sum_k I_k g(t-kT_s),
\end{align} 
where the symbol period is $T_s$ and $\{I_k\}$ is the sequence of amplitudes drawn from a $2^b$ element constellation (e.g. BPSK, 4-PAM, 4-QAM, etc) by mapping $b$-bit binary blocks from an underlying information sequence $\{d_k\}$. The $g(t)$ is the raised-cosine (RC) pulse shaping filter given in \cite[[eq.
4.49]{Stuber02}.
For simplicity, we herein assume BPSK modulation resulting in real amplitudes, which in turn ensures $s(t)$ is a real signal and $b=1$. It is also assumed that the sequence $\{I_k\}$ is normalized such that $\mathcal{E}\left\{|I_k|^2\right\}=1$, which due to the effect of pulse shaping in turn gives $\mathcal{E}\left\{|s(t)|^2\right\}=1-0.25\beta$.
Let the delay spread be defined by \cite{Gold05}:
\begin{align}
	T_m=\max_n |\tau_n-\tau_0|.
\end{align}
In this paper, a mediumband channel is considered, so the symbol period is such that $T_m \leq T_s \leq 10T_m$ \cite{Bas2023}.
\subsection{Computer Simulation Environment}\label{sec:simulation0}
In order to assess the statistics of mediumband channels, and also to compare them with other channel models, the generic propagation model described in \eqref{eq:channel:1} is simulated on MATLAB for NLoS fading scenarios. Typically $\tau_n$ values are dependent on the environment, but without loss of generality, we assume $\tau_0=0$. The other delays $\tau_n$ for $n=1,\dots,N-1$ are drawn from a uniform distribution, $U[0,T_m]$, where $T_m$ is the delay spread. The sequence of amplitudes are drawn from a BPSK constellation, so $\{I_k\} \in \left\{-1, 1\right\}$ for $\forall k$. Furthermore, the phases are drawn from a uniform distribution, $\phi_n \sim U[0,2\pi]$, and two scenarios for path gains namely ``uniform'' and ``exponential'' are considered. In the scenario of ``uniform'', the average path gains are such that $\mathcal{E}\left\{\alpha_1^2\right\}=\mathcal{E}\left\{\alpha_2^2\right\}=\dots =\mathcal{E}\left\{\alpha_n^2\right\} \propto 1/N$. In the ``exponential'' scenario, the average path gains are chosen, for simplicity, such that  $\mathcal{E}\left\{\alpha_n^2\right\} \propto e^{-2\kappa n}$, where $\kappa$ captures the decay of the strength of multipath components. In both scenarios, the path gains are also normalized such that $\mathcal{E}\left\{\sum \alpha_n^2\right\}=1$ \cite{Jakes75}. Furthermore, in both scenarios, $n$th instantaneous path gain is drawn from a Rayleigh distribution, that is $\alpha_n \sim \text{Rayleigh}\left(0.5 \overline{\alpha_n^2}\right)$, where $\overline{\alpha_n^2}=\mathcal{E}\left\{\alpha_n^2\right\}$ is the average gain of the $n$th path. \cite{Molisch14}. 
By changing $T_m$  appropriately while keeping $T_s$ fixed, mediumband channels with different degrees of mediumband-ness are obtained. Unless otherwise is specified, in all simulations in this paper, $T_s=1$ is assumed.
\section{Mediumband Channel Characterization}
The appropriate IO relationship, that captures the effect of mediumband-ness, in the presence of noise, is \cite{Bas2023}:
\begin{align}\label{eq:r(t)}
	r'(t) &= \underbrace{\sqrt{E_s}h_os(t-\hat{\tau})+\sqrt{E_s}\eta_o u(t)}_{r(t)} + n(t),
\end{align}
where $s(t-\hat{\tau})$ is the desired signal; $u(t)$ is a complex uncorrelated zero mean unit variance interference signal; and $\hat{\tau}$ is the time instance, which the RX synchronizes to. Here $n(t)$ is a complex zero mean additive-white-Gaussian-noise (AWGN) signal with a variance of $\sigma^2$. The fading factors $h_o$ and $\eta_o$ are given respectively by: 	
	\begin{align}\label{eq:H}
		h_o &= \frac{\sum_{n=0}^{N-1} \gamma_n R(\tau_n-\hat{\tau})}{1-\frac{\beta}{4}},
	\end{align}
and \eqref{alpha:eq1_opt}, where the complex channel gain,  $\gamma_n=\alpha_ne^{-j\phi_n}$ $\forall n$, and $R(\tau)=\mathcal{E}\left\{s(t)s(t+\tau)\right\}$ is the autocorrelation function of the channel input process $s(t)$, which is given in \eqref{auto-corr:eq1}.\\
\indent In contrast, in the narrowband region, that is when PDS $\leq 10\%$, the distribution of multipath delays are such that, $\tau_0 \approx \tau_1 \dots \approx \tau_{N-1} \approx  \hat{\tau}$. So the received baseband equivalent signal in \eqref{eq:channel:1}  reduces to \cite{Gold05}:
\begin{align}\label{eq:channel:20}
	\setcounter{equation}{8}
	r(t) &\approx \sqrt{E_s} \left(\sum_{n=0}^{N-1} \gamma_n \right) s(t-\hat{\tau}) = \sqrt{E_s} g_o s(t-\hat{\tau})
\end{align}
where $g_o=\sum_{n=0}^{N-1} \gamma_n$. Here the multipath components are said to be nonresolvable, and combined to a single multipath with delay $\hat{\tau}$. The symbol timing synchronizer at the RX typically synchronizes to this common delay, $\hat{\tau}$ \cite{Gold05}. Typically, as $\textup{PDS}$ decreases, it has been shown that $h_o \rightarrow g_o$ and $\eta_o \rightarrow 0$. So, the model in \eqref{eq:r(t)} is general, and can accommodate both narrowband and mediumband scenarios, if needed.
\subsection{Discrete-Time Mediumband Channel}
If the timing recovery at the RX is assumed to be perfect, the discrete-time version of \eqref{eq:r(t)} becomes:
\begin{align}\label{channel:DT:eq1}
	r'(k) &= \overbrace{\sqrt{E_s}h_os(k)+\underbrace{\sqrt{E_s} \eta_o u(k)}_{\mathcal{I}(k)}}^{r(k)} + n(k), 
\end{align}
where it is assumed that $r'(t)$ is sampled regularly at $t=\hat{\tau}+kT_s$. Let the average received signal-to-noise-ratio (SNR) be defined by:
\begin{align}
	\text{Average Received SNR:  }	\bar{\gamma} &= \frac{\mathcal{E}\left\{\left|r(k)\right|^2\right\}}{\mathcal{E}\left\{\left|n(k)\right|^2\right\}}.
\end{align}
where the expectations in both numerator and the denominator are over all propagation parameters and the input process, $s(t)$.  
\subsection{Probability Density Function of $h_o$}\label{sec_pdf_of_h}
\begin{figure}[t]
	\centerline{\includegraphics[scale=0.65]{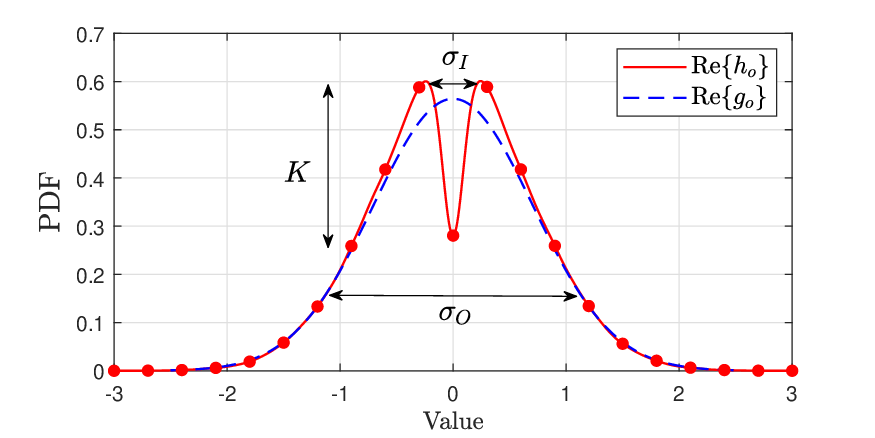}}
	\caption{A typical probability density function of the desired fading factor (i.e., $h_o$) in mediumband channels. The PDF of the desired fading factor of the ideal narrowband channel, i.e. $g_o$ in \eqref{eq:channel:20}, is also shown for comparison. Note that it is the same NLoS environment, which gives rise to the BLUE PDF for $g_o$, that gives rise to the RED PDF for $h_o$. It is the different operating points on the $T_mT_s$--plane that create this difference, but not different environments.}\label{fig:secv:fig1}
\end{figure}

The mediumband channel is a battleground of two opposing phenomena. As PDS increases, or in other words, as the degree of mediumbandness increases, the strength of the additive interference term, $\mathcal{I}(k)$ increases, which is undesirable. However, as PDS increases, the probability of deep fading in $h_o$ decreases rapidly as well. The Fig. \ref{fig:secv:fig1} depicts the probability density function (PDF) of $g_o$ and $h_o$. Here $g_o$ corresponds to the narrowband channel in \eqref{eq:channel:20}, or $\textup{PDS}=0\%$. It is Gaussian distributed and unimodal, and owing to the peak, it can be seen clearly that the probability of $g_o$ being in deep fade is high. In contrast, PDF of $h_o$ has a bimodal distribution, and notably owing to the dip in the middle, the probability of $h_o$ being in deep fade appears to be relatively low. This particular bimodal nature can be accurately modelled using three primary parameters, $\sigma_O$, $\sigma_I$ and $K$, where $\sigma_I$ and $K$ capture the width and depth of the trench in PDF of $h_o$ respectively \cite{BasJ23}:
\begin{subequations}\label{pdfs:eq1}
	\begin{align}
		f_{h_o^i}(x) &= \frac{e^{-\frac{x^2}{2\lambda_0^2}}-Ke^{-\frac{x^2}{2 \lambda_1^2}}}{\sqrt{2\pi \lambda_0^2} - K \sqrt{2\pi \lambda_1^2}},
	\end{align}
	\vspace{-4mm}
	\begin{align}
		f_{h_o^q}(y) &= \frac{e^{-\frac{y^2}{2\lambda_0^2}}-Ke^{-\frac{y^2}{2 \lambda_1^2}}}{\sqrt{2\pi \lambda_0^2} - K \sqrt{2\pi \lambda_1^2}}, 
	\end{align}
\end{subequations}
where $\lambda_0=\sigma_O$ and $\lambda_1=\sqrt{\frac{\sigma_O^2\sigma_I^2}{\sigma_O^2+\sigma_I^2}}$. Consequently,  $f_{h_o}(x,y)=f_{h_o^i}(x)f_{h_o^q}(y)$. Note that $h_o^i$ and $h_o^q$ denote the real and imaginary parts of $h_o$. It can be seen that as $K,\sigma_I \rightarrow 0$, $f_{h_o^i}(x)$ and $	f_{h_o^q}(y)$ collapse to Gaussian distributions. Also:
\begin{align}
	\mathcal{E}\left\{h_o\right\} &= 0,\\
	\mathcal{E}\left\{\left|h_o\right|^2\right\} &= 2 \left(\frac{\lambda_0^3-K\lambda_1^3}{\lambda_0-K\lambda_1}\right).
\end{align}
Typical values for $K,\sigma_O$ and $\sigma_I$ for different PDSs are tabulated in Table \ref{table:1}, whereas $\mathcal{E}\left\{g_o\right\}=0$ and $\mathcal{E}\left\{\left|g_o\right|^2\right\}=1$.   
\begin{table}
	\begin{center}
		\caption{Parameters for PDF of $h_o$ for Different PDSs, where $\kappa=0$.}
		\label{table:1}
		\begin{tabular}{|c|c|c|c|} 
			\hline
			PDS $[\%]$ & $K$  & $\sigma_I^2$ & $\sigma_O^2$\\
			\hline
			0 & 0 & 0 & 0.5000\\
			\hline 
			20 & 0.470 & 0.0045 & 0.4500\\
			\hline
			40 & 0.660 & 0.0090 & 0.4310\\
			\hline  
			60 & 0.770 & 0.0200 & 0.4200\\ 
			\hline
			80 & 0.830 & 0.0280 & 0.4160\\ 
			\hline
		\end{tabular}
	\end{center}
\end{table}
\section{Average Bit Error Rate Performance}\label{sec_bit_error}
%
%
The optimum modulation/detection for, and the corresponding optimum bit error rate (BER) performance of the mediumband channel in \eqref{channel:DT:eq1} is still an open problem. However, in light of \eqref{pdfs:eq1}, a lower bound for BER can be derived. 
\subsection{Analytical Lower Bound}\label{sec_bit_error_bound}
A lower bound for the BER of mediumband channels can be derived by assuming the additive ISI, $\mathcal{I}(k)$ is zero in \eqref{channel:DT:eq1}. As a result, in conjunction with the assumption of perfect knowledge of $h_o$ at the RX, the channel in \eqref{channel:DT:eq1} reduces to a discrete-time AWGN channel, where the optimum modulation/detection is known. So, the average BER of digital modulations with coherent detection $P_e$ can be lower bounded by $P_e \geq P_e^{\textup{LB}}$, where:
\begin{align}\label{ber:eq1}
	P_e^{\textup{LB}} &= \mathcal{E}_{h_o}\left\{\rho_1 Q\left(\sqrt{\rho_2 \bar{\gamma}\left|h_o\right|^2}\right)\right\},
\end{align}
where $Q(.)$ is the standard Q-function and the parameters $\rho_1$ and $\rho_2$ are dependent on the modulation scheme. For instance, for BPSK, $\rho_1=1$ and $\rho_2=2$ \cite{Gold05}. In light of the PDF in \eqref{pdfs:eq1}, \eqref{ber:eq1} can be evaluated as:   
\begin{Corollary}\label{corollary1}
The average BER $P_e$ of mediumband channels with coherent detection can be lower bounded by:
\begin{align}\label{corr1:eq1}
	P_e^{\textup{LB}} &= \frac{\rho_1}{\pi} \int_0^{\pi/2} \left[\frac{\xi\sin \theta}{\sqrt{\rho_2 \lambda_0^2 \bar{\gamma}+\sin ^2\theta}}\right. \qquad \qquad \quad \\ \nonumber
	& \qquad \qquad \qquad \qquad \left. -\frac{(\xi-1)\sin\theta}{\sqrt{\rho_2 \lambda_1^2 \bar{\gamma}+\sin ^2\theta}}\right]^2 d\theta,
\end{align}
where $\xi=\frac{\lambda_0}{\lambda_0-K\lambda_1}$.
\end{Corollary}
\begin{proof}
	A concise proof is available in Appendix \ref{Appendix:A}.
\end{proof}   
\vspace{0mm}
\subsection{Diversity Order Analysis}\label{sec_diversity}
As $\bar{\gamma} \rightarrow \infty$, $P_e^{\textup{LB}}$ can be tightly approximated by:
\begin{align}\label{corr1:eq2}
	P_e^{\textup{LB}} &\approx \left[\frac{\rho_1}{4\rho_2} \left(\frac{\xi}{\lambda_0} - \frac{\xi-1}{\lambda_1}\right)^2 \right] \bar{\gamma}^{-1},
\end{align}
which shows that the diversity order in $P_e^{\textup{LB}}$ is unity. Given the fact that system model herein is a single-input-single-output channel, this result is in agreement with the analysis in \cite{ZhGi03, ZhTse03}. However, the situation in the mediumband region is not as straightforward as equation \eqref{corr1:eq2} portraits. Owing to the higher number of degrees of freedom in the PDF of $h_o$ (see Fig. \ref{fig:secv:fig1}), characterizing error probability in the mediumband regime appears to be complex. We firstly report some emblematic BER results using computer simulation.\\
\indent In the absence of an optimum detection scheme for mediumband channels, we herein adopt two suboptimal methods. Based on \eqref{channel:DT:eq1}, in the first method, the sufficient statistics for the detection of the $k$th information amplitude, $I_k$ is:
\begin{align}
&\!\!\!\!\!\!\!\!\!\!\!\!\!\!\!\!\!\!\!\!\!\!\!\!\!\!\!\!\!\!\!\!\!\!\!\!\!\!\!\!\!\!\!\!\!\!\!\!\!\!\!\!\!\!\!\!\!\textup{\textbf{Method 1:}} \nonumber \\
\Delta(k) &= \textup{Re}\left[\frac{h_o^Hr'(k)}{\sqrt{E_s}|h_o|^2}\right],
\end{align}
and in the 2nd method, the additive interference term, $\sqrt{E_s}\eta_o u(t)$ in \eqref{eq:r(t)} is resolved with two additional taps. So, the received baseband signal is modelled by symbol-period apart three taps as \cite[Corollary 1]{Bas2023}:
\begin{align}\label{appA:gmb:eq5}
\!\!\!r(t) &=\!\sqrt{E_s}\left(h_1\right)_o s(t-\hat{\tau})  + \nonumber \\ 
	&\sum_{v=2}^3 \!\sqrt{E_s}\left(h_v\right)_o s(t-\hat{\tau}-(v\!-\!1)T_s)\!+\!\sqrt{E_s}\zeta_o \bar{u}(t),
\end{align}
where $\sqrt{E_s}\zeta_o \bar{u}(t)$ the residual ISI. If these taps are known at the RX, successive interference cancellation (SIC) can be formulated as:
\begin{align}
&\!\!\!\!\!\!\!\!\!\!\!\!\!\textup{\textbf{Method 2:}} \nonumber \\
\!\!\!\Delta(1)\!&=\! \textup{Re}\!\!\left[\frac{\left(h_1\right)_o^Hr'(1)}{\sqrt{E_s}|\left(h_1\right)_o|^2}\right],\\
\!\!\!\Delta(2)\!&=\! \textup{Re}\!\!\left[\frac{\left(h_1\right)_o^H\left(r'(2)-\left(h_2\right)_o \hat{I}_1\right)}{\sqrt{E_s}|\left(h_1\right)_o|^2}\right],\\
\!\!\!\Delta(k)\!&=\! \textup{Re}\!\!\left[\frac{\left(h_1\right)_o^H\!\left(r'(k)-\left(h_2\right)_o \hat{I}_{k-1} -\left(h_3\right)_o \hat{I}_{k-2} \right)}{\sqrt{E_s}|\left(h_1\right)_o|^2}\right]
\end{align}
for $k=3,4,\dots, 100$, where the frame length is 100bits. In both methods, which adopt symbol-by-symbol detection, the $k$th detected BPSK symbol hence is:
\begin{align} \label{nb:statistics:eq3}
	\hat{I}_k &= \begin{cases}
		1, & \Delta(k) > 0, \\
		-1, & \Delta(k) < 0.
	\end{cases} \qquad \qquad \forall k.
\end{align}   
\section{Simulation Study}\label{simulation_study}
Figs. \ref{fig:fig1} and \ref{fig:fig2} show the BER performance from ``Method 1'' and ``Method 2'' alongside the lower bound in Corollary \ref{corollary1} for BPSK modulation for two emblematic propagation scenarios. In the first scenario, the path gain profile is exponential ($\kappa=0.25$) and PDS$=20\%$. In the second scenario, the path gain profile is uniform (i.e., $\kappa=0$) and PDS$=60\%$. Both figures include the BER performance of the narrowband channel in \eqref{eq:channel:20} with no ISI for comparison (see BLUE curve), which is:
\begin{align}\label{BER:Ray:eq1}
	P_e^{\textup{Rayleigh}} &= 0.5 \left(1- \sqrt{\frac{\bar{\gamma}}{1+\bar{\gamma}}}\right),
\end{align}
and is known to have a unity diversity order\cite{ZhTse03}. This is a widely studied performance limit for NLoS environments. 
\subsubsection*{\textbf{Scenario 1}} This is a more favourable channel for wireless communication, because $\kappa$ is large, which means most of the received power is concentrated in a few multipath components. Also the small dip in Fig. \ref{fig:fig3} for PDS$=20\%$ means the effect of deep fading avoidance in this scenario is small. Meanwhile, Fig. \ref{fig:fig1} shows the lower bound in Corollary \ref{corollary1} along with average BER of Method 1 and 2 for BPSK modulation. Clearly the lower bound falls at a faster rate than that of \eqref{BER:Ray:eq1}, which in fact achieve a diversity order of two. Despite having ISI, average BER of Method 1 outperforms the narrowband case, in some cases by 2.5dB, upto SNR of 40dB. The average BER of Method 2 outperforms the narrowband case with a more significant margin (some cases by over 6.5dB), and is closer to the lower bound. These suggest that the lower bound in \eqref{corr1:eq1} is something that is approachable with reasonably complicated detection methods.
\subsubsection*{\textbf{Scenario 2}} 
\begin{figure}[t]
	\centerline{\includegraphics[scale=0.65]{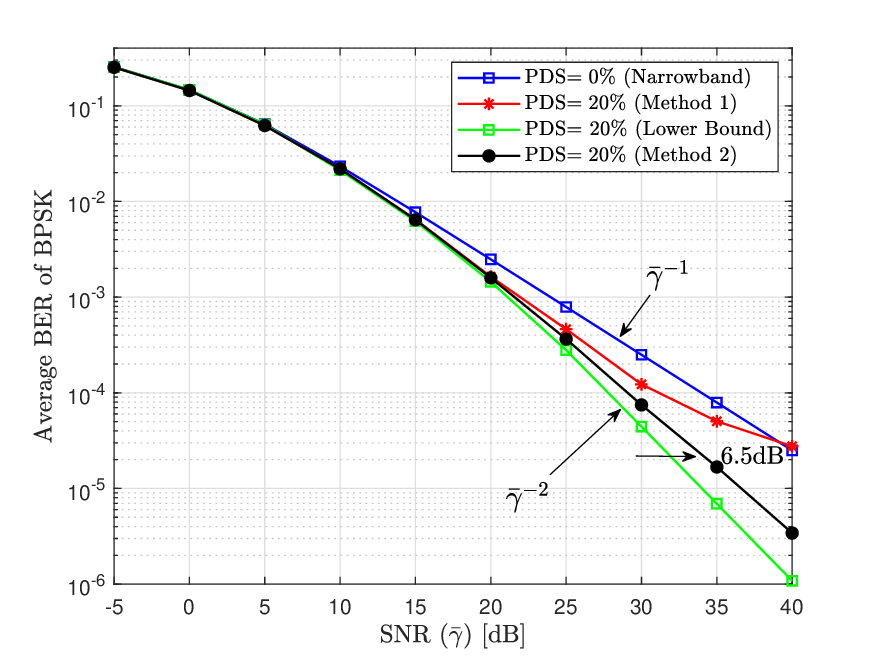}}
	\caption{Average BER of BPSK modulation in a mediumband channel with $\kappa=0.25$ (i.e., Scenario 1), $N=10$, and the roll-off factor $\beta=0.22$.}\label{fig:fig1}
\end{figure}

\begin{figure}[t]
	\centerline{\includegraphics[scale=0.65]{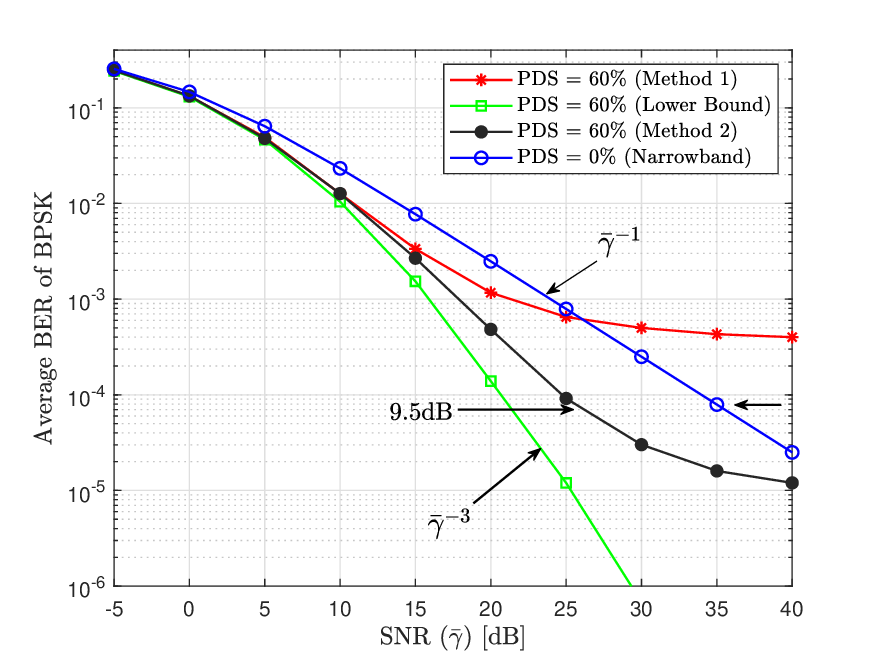}}
	\caption{Average BER of BPSK modulation in a mediumband channel with $\kappa=0$ (i.e., Scenario 2), $N=10$, and $\beta=0.22$.}\label{fig:fig2}
\end{figure}

\begin{figure}[t]
	\centerline{\includegraphics[scale=0.65]{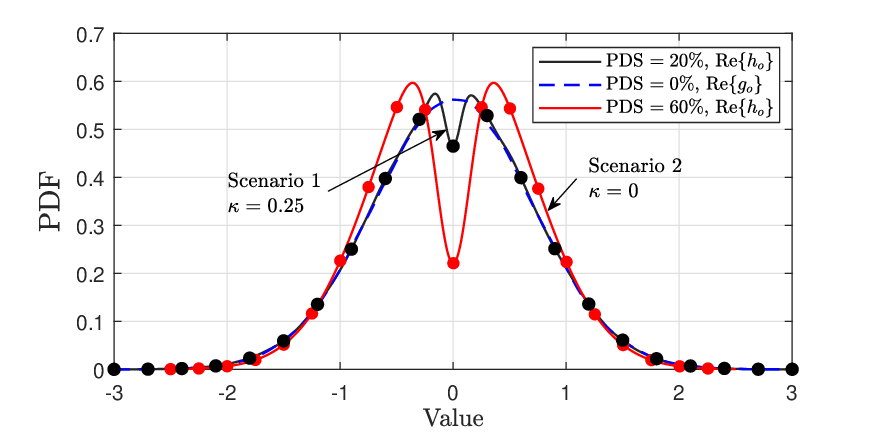}}
	\caption{PDF of $h_o$ (in \eqref{eq:H}) for the two NLoS scenarios considered in Sec. \ref{simulation_study} along with the PDF of $g_o$ (in \eqref{eq:channel:20}).}\label{fig:fig3}
\end{figure}

This is a more hostile channel for wireless communication, because the delay spread is large, and on average, the multipath components are equally strong (i.e, $\kappa=0$). The amount of ISI is high in this channel, but on the other hand, the probability of deep fading is significantly low (See. Fig. \ref{fig:fig3}) too. However, the lower bound now drops at a even faster rate, in fact with a diversity order of three. The average BERs of Method 1 and 2 follow the lower bound upto SNR of 15dB, and outperform the narrowband channel upto SNR of 25 and 40 dB respectively. At BER$=0.8e\text{-4}$, Method 2 outperforms the narrowband channel by about 9.5dB. The following proposition sheds light on trends in Figs. \ref{fig:fig1} and \ref{fig:fig2}.     
\begin{Proposition}\label{proposition2} The lower bound for the average BER of mediumband channels can be expressed as a convergent series as:  
\begin{align}
	P_e^{\textup{LB}} &= \Gamma_1 \bar{\gamma}^{-1} + \Gamma_2 \bar{\gamma}^{-2} + \Gamma_3 \bar{\gamma}^{-3} + \dots
\end{align}
where $\Gamma_1=\frac{\rho_1\left(1-K\right)^2}{4\rho_2\left(\lambda_0-K\lambda_1\right)^2}$ and:
%
%
%
\begin{align}
\Gamma_2&=\frac{-3\rho_1\left(1-K\right)}{16\rho_2^2\left(\lambda_0-K\lambda_1\right)^2}\!\!\left(\frac{1}{\lambda_0^2}-\frac{K}{\lambda_1^2}\right)
\end{align}
\begin{align}
\Gamma_3 &=\frac{5\rho_1}{128\rho_2^3\left(\lambda_0-K\lambda_1\right)^2}\!\!\left[\left(\frac{1}{\lambda_0^2}-\frac{K}{\lambda_1^2}\right)^2 \right. \nonumber \\
&\qquad \qquad \qquad \qquad \qquad \left. +3(1-K)\left(\frac{1}{\lambda_0^4}-\frac{K}{\lambda_1^4}\right)\right]
\end{align}	
\end{Proposition}   
\begin{proof}
	A concise proof is available in Appendix \ref{Appendix:B}.
\end{proof}
We can see that in the narrowband region, that is when $K,\sigma_I^2 \rightarrow 0$, all $\Gamma_1$, $\Gamma_2$ and $\Gamma_3$ are in the same order of magnitude (0.25,0.1875 and 0.1562 respectively), but as PDS increases, that is when $K \rightarrow 1$, $\Gamma_1$ is seen decreasing rapidly. As a result, it is the $\Gamma_2$ and $\Gamma_3$ that are dominant in mid to high PDS regions explaining the higher-than-one order diversity in average BER as PDS increases. However, the average BER fall-off rate ultimately reduces to one as shown in \eqref{corr1:eq2}, but that happens in practically irrelevant ultra high SNR regions.
\begin{figure*}
	\begin{align}\label{eq:Pe:AppB:4}
		\setcounter{equation}{37}
		P_e^{\textup{LB}} &\approx \frac{\rho_1}{\pi} \int_0^{\pi/2} \left[\xi^2\bar{x}-2\xi\left(\xi-1\right)\sqrt{\bar{x}\bar{y}} + (\xi-1)^2\bar{y}\right] d\theta \qquad \qquad \quad \\ \nonumber
		& \qquad \qquad \qquad \qquad \qquad
		- \frac{\rho_1}{\pi} \int_0^{\pi/2} \left[\xi^2\bar{x}^2 - 2\xi(\xi-1)\sqrt{\bar{x}\bar{y}}\left(\frac{\bar{x}}{2}-\frac{\bar{y}}{2}\right) + (\xi-1)^2\bar{x}^2\right] d\theta \qquad \qquad \quad \\ \nonumber
		& \qquad \qquad \qquad \qquad 
		+ \frac{\rho_1}{4\pi} \int_0^{\pi/2} \left[4\xi^2\bar{x}^3 - 2\xi\left(\xi-1\right)\bar{x}^{\frac{3}{2}}\bar{y}^{\frac{3}{2}} - 3\xi\left(\xi-1\right)\bar{x}^{\frac{1}{2}}\bar{y}^{\frac{5}{2}} - 3\xi\left(\xi-1\right)\bar{x}^{\frac{5}{2}}\bar{y}^{\frac{1}{2}} + 4(\xi-1)^2\bar{y}^3
		\right] d\theta
	\end{align}
\hrule
\end{figure*} 
%
\section{Conclusion}
This paper studied the reliability of mediumband wireless communication in terms of the average BER. The results, backed by statistical analysis and computer simulations, have shown that mediumband channels can be designed to achieve significantly superior BER and diversity, outperforming some key fundamental limits, due mainly to the effect of deep fading avoidance. The gains, offered by mediumband channels even with suboptimal detection schemes, have been shown to be many dBs. It is anticipated that more sophisticated detection methods (e.g. ML based) would approach the performance limits (i.e., Corollary \ref{corollary1}) predicted herein. 
\appendices
\section{Proof of Corollary \ref{corollary1}}\label{Appendix:A}
\begin{align}\label{ber:corr:eq1}
P_e^{\textup{LB}} &= \rho_1 \int\limits_{-\infty}^{\infty} \int\limits_{-\infty}^{\infty}
	Q\left(\sqrt{\rho_2\bar{\gamma}\left(x^2+y^2\right)}\right) f_{h_o}(x,y) dx dy, \nonumber
\end{align}
where $f_{h_o}(x,y)$ is the joint PDF of the real and imaginary parts of $h_o$, which can be found in \eqref{pdfs:eq1}. In light of \cite{Craig91}, expressing $Q(.)$ in integral form $P_e^{\textup{LB}}$ can be rewritten as:
\begin{align}
P_e^{\textup{LB}} &= \dfrac{\rho_1}{\pi} \int\limits_{-\infty}^{\infty} \int\limits_{-\infty}^{\infty} \int_0^{\frac{\pi}{2}} e^{\frac{-\rho_2\bar{\gamma}\left(x^2+y^2\right)}{2\sin^2 \theta}} f_{h_o}(x,y) d\theta dx dy, 
\end{align}
which can be rewritten, due to the finite limits of integrals, as:
\begin{align}\label{ber:corr:eq3}
P_e^{\textup{LB}} &=\frac{\rho_1}{\pi} \int_0^{\frac{\pi}{2}} \int\limits_{-\infty}^{\infty} \int\limits_{-\infty}^{\infty}  e^{\frac{-\rho_2\bar{\gamma}\left(x^2+y^2\right)}{2\sin^2 \theta}}f_{h_o}(x,y) dx dy d\theta.  
\end{align}
The fact $f_{h_o}(x,y)=f_{h_o^i}(x)f_{h_o^q}(y)$ and the interchangeability of $x$ and $y$ in \eqref{ber:corr:eq3} yield:
\begin{align}\label{ber:corr:eq4}
	\!\!\!P_e^{\textup{LB}} &= \dfrac{\rho_1}{\pi} \int_0^{\frac{\pi}{2}} \left[\int\limits_{-\infty}^{\infty}  e^{-\left(\frac{\rho_2\bar{\gamma}x^2}{2\sin^2\theta}\right)}f_{h_o^i}(x)dx\right]^2 d\theta,
\end{align}
which can be elaborated as:
\begin{align}\label{ber:corr:eq5}
	\!\!\! P_e^{\textup{LB}} &= \dfrac{\rho_1}{\pi} \int_0^{\frac{\pi}{2}} \left[\bar{\alpha}\int \limits_{-\infty}^{\infty}  e^{-\left(\frac{\rho_2\bar{\gamma}}{2\sin^2\theta}+\frac{1}{2\lambda_0^2}\right)x^2}dx \right. \\
	&\qquad \qquad \qquad -\left. \bar{\beta} \int\limits_{-\infty}^{\infty}  e^{-\left(\frac{\rho_2\bar{\gamma}}{2\sin^2\theta}+\frac{1}{2\lambda_1^2}\right)x^2} dx \right]^2 d\theta,
\end{align}
where $\bar{\alpha}=\frac{1}{\sqrt{2\pi \lambda_0^2} - K\sqrt{2\pi\lambda_1^2}}$ and $\bar{\beta}=\frac{K}{\sqrt{2\pi \lambda_0^2} - K\sqrt{2\pi\lambda_1^2}}$. Using the integral identity:
\begin{align}
	\int\limits_{-\infty}^{\infty} e^{-ax^2}dx &= \sqrt{\frac{\pi}{a}},
\end{align}
followed by some algebraic manipulations lead to the final result as shown in \eqref{corr1:eq1}.
\section{Proof of Proposition \ref{proposition2}}\label{Appendix:B}
From Corollary 1, a lower bound for average BER is:
\begin{align}\label{eq:Pe:AppB:1}
	\setcounter{equation}{34}
	P_e^{\textup{LB}} &= \frac{\rho_1}{\pi} \int_0^{\pi/2} \left[\frac{\xi\sin \theta}{\sqrt{\rho_2 \lambda_0^2 \bar{\gamma}+\sin ^2\theta}}\right. \qquad \qquad \quad \\ \nonumber
	& \qquad \qquad \qquad \qquad \qquad\left. -\frac{(\xi-1)\sin\theta}{\sqrt{\rho_2 \lambda_1^2 \bar{\gamma}+\sin ^2\theta}}\right]^2 d\theta,
\end{align}
which can also be expressed by:
\begin{align}\label{eq:Pe:AppB:2}
	P_e^{\textup{LB}} &= \frac{\rho_1}{\pi} \int_0^{\pi/2} \left[\frac{\xi\sqrt{\bar{x}}}{\sqrt{1+\bar{x}}} -\frac{(\xi-1)\sqrt{\bar{y}}}{\sqrt{1+\bar{y}}}\right]^2 d\theta,
\end{align}
where $\bar{x}=\sin^2\theta/\rho_2 \lambda_0^2 \bar{\gamma}$ and $\bar{y}=\sin^2\theta/\rho_2 \lambda_1^2 \bar{\gamma}$. It is clear that both $\bar{x}$ and $\bar{y}$ are less than unity as $\bar{\gamma}\rightarrow \infty$. So using the Taylor expansion only for the denominator terms:  $\left(1+\bar{x}\right)^{-\frac{1}{2}}\approx 1-\frac{1}{2}\bar{x}+\frac{3}{8}\bar{x}^2$ and $\left(1+\bar{y}\right)^{-\frac{1}{2}}\approx 1-\frac{1}{2}\bar{y}+\frac{3}{8}\bar{y}^2$, one can arrive at:
\begin{align}\label{eq:Pe:AppB:3}
	P_e^{\textup{LB}} &\approx \frac{\rho_1}{\pi} \int_0^{\pi/2} \left[\xi\sqrt{\bar{x}}\left(1-\frac{1}{2}\bar{x}+\frac{3}{8}\bar{x}^2\right) \right. \qquad \qquad \quad \\ \nonumber
	& \qquad \qquad \qquad \left. - (\xi-1)\sqrt{\bar{y}}\left(1-\frac{1}{2}\bar{y}+\frac{3}{8}\bar{y}^2\right)\right]^2 d\theta.
\end{align}
The expression \eqref{eq:Pe:AppB:3} can be further expanded as in \eqref{eq:Pe:AppB:4}. In light of the fact that $\xi=\frac{\lambda_0}{\lambda_0-K\lambda_1}$, $P_e^{\textup{LB}}$ in \eqref{eq:Pe:AppB:4} can be simplified to get Proposition \ref{proposition2}, where $\int_o^{\pi/2} \sin^n \theta d \theta=0.5B\left(\frac{n+1}{2},\frac{1}{2}\right)$ is used. Here $B(.,.)$ is the standard Beta function.        

\begin{thebibliography}{00}
%
%
\bibitem{Zhang23} 
Z. Zhang et al., ``Active RIS vs. passive RIS: Which will prevail in 6G?,'' IEEE Transactions on Commun., vol. 71, no. 3, pp. 1707--1725, Mar. 2023.
%
\bibitem{Mei22} 
W. Mei, B. Zheng, C. You and R. Zhang, ``Intelligent reflecting surface-aided wireless networks: From single-reflection to multireflection design and optimization,'' Proceedings of the IEEE, vol. 110, no. 9, pp. 1380--1400, Sep. 2022.
%
\bibitem{Bas2023}
D. A. Basnayaka, ``Introduction to mediumband wireless communication,'' IEEE Open Journal of the Commun. Society, vol. 4, pp. 1247--1262, May. 2023.
%
\bibitem{BasJ23}
D. A. Basnayaka and P. J. Smith, ``The effect of deep fading avoidance in mediumband radio frequency channels,'' Proceedings of IEEE VTC-Fall, Hong-Kong, Oct. 2023.
%
%
\bibitem{ZhGi03}
Z. Wang and G. B. Giannakis, ``A simple and general parameterization quantifying performance in fading channels,'' IEEE Transactions on Communications, vol. 51, no. 8, pp. 1389--1398, Aug. 2003.
%
\bibitem{ZhTse03}
L. Zheng and D. N. C. Tse, ``Diversity and multiplexing: a fundamental tradeoff in multiple-antenna channels,'' IEEE Trans. on Info. Theory, vol. 49, no. 5, pp. 1073--1096, May 2003
%
\bibitem{Stuber02}
G. L. Stuber, ``Principals of Mobile Communications,'' 2nd ed., Kluwer Academic Publishers, 2002.
%
%
\bibitem{Molisch14}
A. Meijerink, et. al., ``On the physical interpretation of the Saleh–Valenzuela model and the definition of its power delay profiles,'' IEEE Trans. on Anten. and Propa., vol. 62, no. 9, pp. 4780--4793, Sep. 2014.
%
\bibitem{Gold05}
A. Goldsmith, Wireless Communications, Cambridge University Press,
2005.
%
%
\bibitem{Proakis00}
J. G. Proakis, “Digital Communication,” McGraw-Hill Education, 4th edition, 2000.
%
%
\bibitem{Jakes75}
W. C. Jakes, ``Microwave mobile communications,'', 1st ed., John Wiley \& Sons Inc, New York, 1975.
%
%
\bibitem{Craig91}
W. Craig, ``A new, simple and exact result for calculating the probability of error for two-dimensional signal constellations,'' MILCOM 91 - Conference record, McLean, VA, USA, 1991.
%
\end{thebibliography}
\end{document}